\documentclass{JAC2000}
\usepackage{graphicx}
\newcommand{\ud}{\mathrm{d}}
\begin{document}
\title{MULTISCALE  ANALYSIS OF RMS ENVELOPE DYNAMICS}
\author{A. Fedorova, M. Zeitlin, IPME, RAS, St.~Petersburg, 
V.O. Bolshoj pr., 61, 199178, Russia
\thanks{e-mail: zeitlin@math.ipme.ru}\thanks{http://www.ipme.ru/zeitlin.html;
http://www.ipme.nw.ru/zeitlin.html}}

\maketitle

\begin{abstract}
We present applications of variational -- wavelet approach to 
different forms of nonlinear (rational)
rms envelope equations.
We have the representation for beam bunch oscillations as
a multiresolution (multiscales) expansion in the base of compactly
supported wavelet bases.
\end{abstract}

\section{INTRODUCTION}
In this paper we consider the applications of a new nu\-me\-ri\-cal\--analytical 
technique which is based on the methods of local nonlinear harmonic
analysis or wavelet analysis to the nonlinear 
root-mean-square (rms) envelope dynamics [1].
Such approach may be useful in all models in which  it is 
possible and reasonable to reduce all complicated problems related with 
statistical distributions to the problems described 
by systems of nonlinear ordinary/partial differential 
equations. In this paper we  consider an approach based on 
the second moments of the distribution functions for  the calculation
of evolution of rms envelope of a beam.
The rms envelope equations are the most useful for analysis of the 
beam self--forces (space--charge) effects and also 
allow to consider  both transverse and longitudinal dynamics of
space-charge-dominated relativistic high--brightness
axisymmetric/asymmetric beams, which under short laser pulse--driven
radio-frequency photoinjectors have fast transition from nonrelativistic
to relativistic regime [1]. Analysis of halo growth in beams, appeared
as result of bunch oscillations in the particle-core model, also are based
on three-dimensional envelope equations [2].
From the formal point of 
view we may consider rms envelope equations 
after straightforward transformations to standard Cauchy form 
as a system of nonlinear differential equations 
which are not more than rational (in dynamical variables). 
Because of rational type of
nonlinearities we need to consider some extension of our results from 
[3]-[10], which are based on application of wavelet analysis technique to 
variational formulation of initial nonlinear problems.  
Wavelet analysis is a relatively novel set of mathematical
methods, which gives us a possibility to work with well-localized bases in
functional spaces and give for the general type of operators (differential,
integral, pseudodifferential) in such bases the maximum sparse forms. 
Our approach in this paper is based on the generalization [11] of 
variational-wavelet approach from [3]-[10],
which allows us to consider not only polynomial but rational type of 
nonlinearities. 

Our representation for solution has the following form
\begin{equation}\label{eq:z}
z(t)=z_N^{slow}(t)+\sum_{j\geq N}z_j(\omega_jt), \quad \omega_j\sim 2^j
\end{equation}
which corresponds to the full multiresolution expansion in all time
scales.
Formula (\ref{eq:z}) gives us expansion into a slow part $z_N^{slow}$
and fast oscillating parts for arbitrary N. So, we may move
from coarse scales of resolution to the
finest one for obtaining more detailed information about our dynamical process.
The first term in the RHS of equation (1) corresponds on the global level
of function space decomposition to  resolution space and the second one
to detail space. In this way we give contribution to our full solution
from each scale of resolution or each time scale.
The same is correct for the contribution to power spectral density
(energy spectrum): we can take into account contributions from each
level/scale of resolution.
In part 2 we describe the different forms of rms equations.
In part 3 we present explicit analytical construction for solutions of
rms equations from part 2, which are based on our variational formulation 
of initial dynamical problems and on  multiresolution representation [11].
We give explicit representation for all dynamical variables in the base of
compactly supported wavelets. Our solutions
are parametrized
by solutions of a number of reduced algebraical problems from which one
is nonlinear with the same degree of nonlinearity and the rest  are
the linear problems which correspond to particular
method of calculation of scalar products of functions from wavelet bases
and their derivatives.

\section{RMS EQUATIONS}
Below we consider a number of different forms of RMS envelope equations,
which are from the formal  point of view
not more than nonlinear differential equations with rational
nonlinearities and variable coefficients.
Let $f(x_1,x_2)$ be the distribution function which gives full information
 about 
noninteracting ensemble of beam particles regarding to trace space or 
transverse phase coordinates $(x_1,x_2)$. 
Then we may extract the first nontrivial bit of `dynamical information' from 
the second moments
\begin{eqnarray}
\sigma_{x_1}^2&=&<x_1^2>=\int\int x_1^2 f(x_1,x_2)\ud x_1\ud x_2 \nonumber\\
\sigma_{x_2}^2&=&<x_2^2>=\int\int x_2^2 f(x_1,x_2)\ud x_1\ud x_2 \\
\sigma_{x_1 x_2}^2&=&<x_1 x_2>=\int\int x_1 x_2 f(x_1,x_2)\ud x_1\ud x_2 \nonumber
\end{eqnarray}
RMS emittance ellipse is given by
$
\varepsilon^2_{x,rms}=<x_1^2><x_2^2>-<x_1 x_2>^2
$.
 Expressions for twiss  parameters are also based on the second moments.

We will consider the following particular
cases of rms envelope equations, which described evolution
of the moments (1) ([1],[2] for full designation):
for asymmetric beams we have the system of two envelope equations
of the second order for $\sigma_{x_1}$ and $\sigma_{x_2}$:
\begin{eqnarray}
&&\sigma^{''}_{x_1}+\sigma^{'}_{x_1}\frac{\gamma '}{\gamma}+
\Omega^2_{x_1}\left(\frac{\gamma '}{\gamma}\right)^2\sigma_{x_1}=\\
&&{I}/({I_0(\sigma_{x_1}+\sigma_{x_2})\gamma^3})
+
\varepsilon^2_{nx_1}/{\sigma_{x_1}^3\gamma^2},\nonumber\\
&&\sigma^{''}_{x_2}+\sigma^{'}_{x_2}\frac{\gamma '}{\gamma}+
\Omega^2_{x_2}\left(\frac{\gamma '}{\gamma}\right)^2\sigma_{x_2}=\nonumber\\ 
&&{I}/({I_0(\sigma_{x_1}+\sigma_{x_2})\gamma^3})
+
\varepsilon^2_{nx_2}/{\sigma_{x_2}^3\gamma^2}\nonumber
\end{eqnarray}
The envelope equation for an axisymmetric beam is
a particular case of preceding equations.

Also we have related Lawson's equation for evolution of the rms
envelope in the paraxial limit, which governs evolution of cylindrical
symmetric envelope under external linear focusing channel
of strenghts $K_r$:
\begin{equation}
\sigma^{''}+\sigma^{'}\left(\frac{\gamma '}{\beta^2\gamma}\right)+
K_r\sigma=\frac{k_s}{\sigma\beta^3\gamma^3}+
\frac{\varepsilon^2_n}{\sigma^3\beta^2\gamma^2},\nonumber
\end{equation}
where
$
K_r\equiv -F_r/r\beta^2\gamma mc^2, \ \ \
 \beta\equiv \nu_b/c=\sqrt{1-\gamma^{-2}}
$
According [2] we have the following form for envelope equations in the model of halo formation
by bunch oscillations:
\begin{eqnarray}
\ddot{X}+k_x^2(s)X-\frac{3K}{8}\frac{\xi_x}{YZ}-\frac{\varepsilon^2_x}{X^3}&=&0,\nonumber\\
\ddot{Y}+k_y^2(s)Y-\frac{3K}{8}\frac{\xi_y}{XZ}-\frac{\varepsilon^2_y}{Y^3}&=&0,\\
\ddot{Z}+k_z^2(s)Z-\gamma^2\frac{3K}{8}\frac{\xi_z}{XY}-\frac{\varepsilon^2_z}{Z^3}&=&0,\nonumber
\end{eqnarray}
where X(s), Y(s), Z(s) are bunch envelopes, $\xi_x, \xi_y$, $\xi_z= F(X,Y,Z)$.

After transformations to Cauchy form we can see that
all this equations from the formal point of view are not more than
ordinary differential equations with rational nonlinearities
and variable coefficients
(also,b we may consider regimes in which $\gamma$, $\gamma'$
are not fixed functions/constants but satisfy some additional differential 
constraint/equations,
but this case does not change our general approach).

\section{Rational Dynamics}

Our problems may be formulated as the systems of ordinary differential            
equations                                                              
\begin{eqnarray}\label{eq:pol0}                                
& & Q_i(x)\frac{\ud x_i}{\ud t}=P_i(x,t),\quad x=(x_1,..., x_n),\\
& &i=1,...,n, \quad                                                                        
 \max_i  deg \ P_i=p, \quad \max_i deg \  Q_i=q \nonumber                  
\end{eqnarray}                                                 
with fixed initial conditions $x_i(0)$, where $P_i, Q_i$ are not more    
than polynomial functions of dynamical variables $x_j$                                 
and  have arbitrary dependence of time. Because of time dilation                 
we can consider  only next time interval: $0\leq t\leq 1$.                      
 Let us consider a set of functions                                               
\begin{eqnarray}                                                                      
 \Phi_i(t)=x_i\frac{\ud}{\ud t}(Q_i y_i)+P_iy_i                        
\end{eqnarray}                                                  
and a set of functionals                   
\begin{eqnarray}
F_i(x)=\int_0^1\Phi_i (t)dt-Q_ix_iy_i\mid^1_0,
\end{eqnarray}
where $y_i(t) \ (y_i(0)=0)$ are dual (variational) variables.
It is obvious that the initial system  and the system
\begin{equation}\label{eq:veq}
F_i(x)=0
\end{equation}
are equivalent.
Of course, we consider such $Q_i(x)$ which do not lead to the singular
problem with $Q_i(x)$, when $t=0$ or $t=1$, i.e. $Q_i(x(0)), Q_i(x(1))\neq\infty$.

Now we consider formal expansions for $x_i, y_i$:
\begin{eqnarray}\label{eq:pol1}
x_i(t)=x_i(0)+\sum_k\lambda_i^k\varphi_k(t)\quad
y_j(t)=\sum_r \eta_j^r\varphi_r(t),
\end{eqnarray}
where $\varphi_k(t)$ are useful basis functions of  some functional
space ($L^2, L^p$, Sobolev, etc) corresponding to concrete
problem and
 because of initial conditions we need only $\varphi_k(0)=0$, $r=1,...,N, \quad i=1,...,n,$
\begin{equation}\label{eq:lambda}
\lambda=\{\lambda_i\}=\{\lambda^r_i\}=(\lambda_i^1, \lambda_i^2,...,\lambda_i^N),
\end{equation}
 where the lower index i corresponds to
expansion of dynamical variable with index i, i.e. $x_i$ and the upper index $r$
corresponds to the numbers of terms in the expansion of dynamical variables in the
formal series.
Then we put (\ref{eq:pol1}) into the functional equations (\ref{eq:veq}) and as result
we have the following reduced algebraical system
of equations on the set of unknown coefficients $\lambda_i^k$ of
expansions (\ref{eq:pol1}):
\begin{eqnarray}\label{eq:pol2}
L(Q_{ij},\lambda,\alpha_I)=M(P_{ij},\lambda,\beta_J),
\end{eqnarray}
where operators L and M are algebraization of RHS and LHS of initial problem
(\ref{eq:pol0}), where $\lambda$ (\ref{eq:lambda}) are unknowns of reduced system
of algebraical equations (RSAE)(\ref{eq:pol2}).

$Q_{ij}$ are coefficients (with possible time dependence) of LHS of initial
system of differential equations (\ref{eq:pol0}) and as consequence are coefficients
of RSAE.

 $P_{ij}$ are coefficients (with possible time dependence) of RHS
of initial system of differential equations (\ref{eq:pol0}) and as consequence
are coefficients of RSAE.
$I=(i_1,...,i_{q+2})$, $ J=(j_1,...,j_{p+1})$ are multiindexes, by which are
labelled $\alpha_I$ and $\beta_I$ --- other coefficients of RSAE (\ref{eq:pol2}):
\begin{equation}\label{eq:beta}
\beta_J=\{\beta_{j_1...j_{p+1}}\}=\int\prod_{1\leq j_k\leq p+1}\varphi_{j_k},
\end{equation}
where p is the degree of polinomial operator P (\ref{eq:pol0})
\begin{equation}\label{eq:alpha}
\alpha_I=\{\alpha_{i_1}...\alpha_{i_{q+2}}\}=\sum_{i_1,...,i_{q+2}}\int
\varphi_{i_1}...\dot{\varphi_{i_s}}...\varphi_{i_{q+2}},
\end{equation}
where q is the degree of polynomial operator Q (\ref{eq:pol0}),
$i_\ell=(1,...,q+2)$, $\dot{\varphi_{i_s}}=\ud\varphi_{i_s}/\ud t$.

Now, when we solve RSAE (\ref{eq:pol2}) and determine
unknown coefficients from formal expansion (\ref{eq:pol1}) we therefore
obtain the solution of our initial problem.
It should be noted if we consider only truncated expansion (\ref{eq:pol1}) with N terms
then we have from (\ref{eq:pol2}) the system of $N\times n$ algebraical equations
with degree $\ell=max\{p,q\}$
and the degree of this algebraical system coincides
 with degree of initial differential system.
So, we have the solution of the initial nonlinear
(rational) problem  in the form
\begin{eqnarray}\label{eq:pol3}
x_i(t)=x_i(0)+\sum_{k=1}^N\lambda_i^k X_k(t),
\end{eqnarray}
where coefficients $\lambda_i^k$ are roots of the corresponding
reduced algebraical (polynomial) problem RSAE (\ref{eq:pol2}).
Consequently, we have a parametrization of solution of initial problem
by solution of reduced algebraical problem (\ref{eq:pol2}).
The first main problem is a problem of
 computations of coefficients $\alpha_I$ (\ref{eq:alpha}), $\beta_J$
(\ref{eq:beta}) of reduced algebraical
system.
These problems may be explicitly solved in wavelet approach.
The obtained solutions are given
in the form (\ref{eq:pol3}),
where
$X_k(t)$ are basis functions and
  $\lambda_k^i$ are roots of reduced
 system of equations.  In our case $X_k(t)$
are obtained via multiresolution expansions and represented by
 compactly supported wavelets and $\lambda_k^i$ are the roots of
corresponding general polynomial  system (\ref{eq:pol2}).
Our constructions are based on multiresolution app\-ro\-ach. Because affine
group of translation and dilations is inside the approach, this
method resembles the action of a microscope. We have contribution to
final result from each scale of resolution from the whole
infinite scale of spaces. More exactly, the closed subspace
$V_j (j\in {\bf Z})$ corresponds to  level j of resolution, or to scale j.
We consider  a multiresolution analysis of $L^2 ({\bf R}^n)$
(of course, we may consider any different functional space)
which is a sequence of increasing closed subspaces $V_j$:
$
...V_{-2}\subset V_{-1}\subset V_0\subset V_{1}\subset V_{2}\subset ...
$
satisfying the following properties:
\begin{eqnarray}
&&\displaystyle\bigcap_{j\in{\bf Z}}V_j=0,\quad
\overline{\displaystyle\bigcup_{j\in{\bf Z}}}V_j=L^2({\bf R}^n),\nonumber
\end{eqnarray}
So, on Fig.1 we present contributions to bunch oscillations
from first 5 scales or levels of resolution.
\begin{figure}
\centering                                                                      
\includegraphics*[width=60mm]{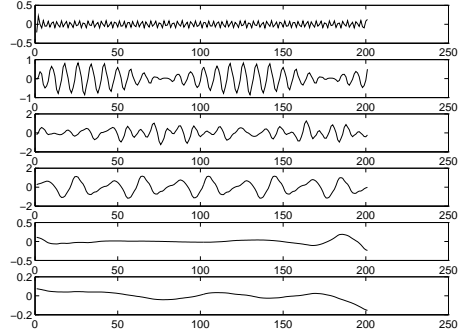}                                     
\caption{Contributions to bunch oscillations: from scale $2^1$ to $2^5$.}           
\end{figure}                                                                           
It should be noted that such representations (1), (15)
for solutions of equations (3)-(5) give the best possible localization
properties in corresponding phase space. This is especially important because 
our dynamical variables corresponds to moments of ensemble of beam particles.

In contrast with different approaches formulae (1), (15) do not use perturbation
technique or linearization procedures 
and represent bunch oscillations via generalized nonlinear localized eigenmodes expansion.

We would like to thank Prof.
J. B. Rosenzweig and Mrs. Melinda Laraneta (UCLA) and Prof. M. Regler (IHEP, Vienna) for
nice hospitality, help and support during
UCLA ICFA Workshop and EPAC00.

\end{document}